\documentstyle[psfig]{mn}
          
\newif\ifAMStwofonts                        
\def\lsimeq{{_<\atop^{\sim}}}
\def\gsimeq{{_>\atop^{\sim}}}

\def\eso36{\hbox{ESO~3.6-m}}                          
\def\nin{\noindent}

\title[ELAIS South 15$\mu$m Source Counts]{A New Method for ISOCAM Data Reduction -- II. 
Mid-Infrared Extragalactic Source Counts in the Southern ELAIS Field}
\author[C. Gruppioni, C. Lari, F. Pozzi et al.]
{\parbox[]{6.5in} {C. Gruppioni$^{1,2}$\thanks{e-mail: gruppioni@pd.astro.it}, C. Lari$^{3}$, 
F. Pozzi$^{2,4}$, G. Zamorani$^{2}$, A. Franceschini$^{5}$, S. Oliver$^{6}$,
M. Rowan-Robinson$^{7}$ and S. Serjeant$^{8}$} \\ \\
$^{1}$ Istituto Nazionale di Astrofisica: Osservatorio Astronomico di Padova, vicolo dell'Osservatorio 5, I--35122 Padova, Italy \\
$^{2}$ Istituto Nazionale di Astrofisica: Osservatorio Astronomico di Bologna, via Ranzani 1, I--40127 Bologna, Italy\\
$^{3}$ Istituto di Radioastronomia del CNR, via Gobetti 101, I--40129 Bologna, Italy\\
$^{4}$ Dipartimento di Astronomia, Universit\`a di Bologna, via Ranzani 1, I--40127 Bologna, Italy\\   
$^{5}$ Dipartimento di Astronomia, Universit\`a di Padova, vicolo dell'Osservatorio 2, I--35122 Padova, Italy\\
$^{6}$ Astronomy Centre, CPES, University of Sussex, Falmer, Brighton BN1 9QJ, UK\\
$^{7}$ Imperial College of Science, Technology and Medicine, Prince Consort Road, London SW7 2BZ, UK\\
$^{8}$ Unit for Space Sciences and Astrophysics, School of Physical Sciences, 
University of Kent, Canterbury, CT2 7NR, UK\\
}

\date{Accepted ??? Received ???}

\pagerange{\pageref{firstpage}--\pageref{lastpage}}

\pubyear{2002}

\def\LaTeX{L\kern-.36em\raise.3ex\hbox{a}\kern-.15em
  T\kern-.1667em\lower.7ex\hbox{E}\kern-.125emX}

\begin{document}

\label{firstpage}

\maketitle                  
      
\begin{abstract}
We present the 15 $\mu$m extragalactic source counts from the Final Analysis Catalogue of 
the European Large Area ISO Survey southern hemisphere field S1, extracted using the {\it Lari 
method}. The large number of extragalactic sources ($\sim 350$) detected over this area 
between about 0.5 and 100 mJy guarantee a high statistical significance of the source 
counts in the previously poorly covered flux density range between IRAS and the Deep 
ISOCAM Surveys. The bright counts in S1 ($S_{15~\mu m} \gsimeq 2$ mJy) are significantly 
lower than other published ISOCAM counts in the same flux range and are consistent with a 
flat, Euclidean slope, suggesting the dominance of a non-evolving population. In contrast,
at fainter 
fluxes ($S_{15~\mu m} \lsimeq 2$ mJy) our counts show a strong departure from 
no evolution models, with a very steep super-Euclidean slope down to our flux limit
($\sim$0.5 mJy). Strong luminosity and density evolution of the order of, respectively, 
$L \propto (1+z)^{3.0}$ and $\rho \propto (1+z)^{3.5}$ is needed at least for the population 
of star-forming galaxies in order to fit the counts and the redshift distributions observed 
at different fluxes. A luminosity break around 10$^{10.8} ~L_{\odot}$ must be introduced in 
the local luminosity function of starburst galaxies in order to reproduce our sharp increase 
of the counts below 2 mJy and the redshift distributions observed for 15 $\mu$m sources
at different flux levels. The contribution of the strongly evolving starburst population 
(down to 50 $\mu$Jy) to the 15 $\mu$m cosmic background is estimated to be $\sim$2.2 nW 
m$^{-2}$ sr$^{-1}$, which corresponds to $\sim$67\% of the total mid-infrared background 
estimate.
\end{abstract}

\begin{keywords}
galaxies: evolution -- galaxies: starburst -- cosmology: observations--infrared: galaxies.
\end{keywords}

\section{Introduction}
Deep galaxy counts are a key instrument for the study of galaxy evolution and can provide strong 
constraints to 
theoretical models. In fact, the departure of source counts from Euclidean predictions depends on the 
intrinsic evolution of galaxies and their redshift distribution. In the past few years several deep observations
in different wave-bands have provided a significant advance in our knowledge of galaxy formation and evolution.  
Optical surveys found a strong evolution of the population of blue galaxies as a function of redshift to $z \sim 1$
(Lilly et al. 1996; Metcalfe et al. 1995), while Mid- and Far-Infrared deep surveys (Elbaz et al. 1999a,b, 2000; 
Dole et al. 1999), 
together with the detection of a substantial diffuse cosmic Infrared Background in the
300 $\mu$m - 1 mm range (Puget et al. 1996; Hauser et al. 
1998; Fixsen et al. 1998; Lagache et al. 1999) implied a strong evolution also for galaxies 
emitting in the infrared.
In fact, the Mid/Far-IR extragalactic background is at least as large as the UV/optical/NIR background, thus
implying a stronger contribution of obscured star formation at redshifts larger than those observed by IRAS.

\nin IRAS has sampled the local Universe ($z < 0.2$) in the Mid/Far-IR band, discovering 
Luminous Infrared Galaxies (LIGs: $L > 10^{11}L_{\odot}$) which radiate most of their light 
in the infrared band. Although LIGs are the most luminous starburst galaxies ever 
detected, they are relatively rare in the local Universe, thus making up only a small 
fraction of the total energy output from galaxies (Soifer \& Negeubauer 1991). However, 
different analyses of the IRAS extragalactic source counts by Hacking, 
Houck and Condon (1987), Lonsdale \& Hacking (1989), Saunders et al. (1990) and Kim and 
Sanders (1998) have shown some evidence for strong evolution at low flux density levels for 
ULIGs (Ultra-LIGs: $L_{Bol} \simeq L_{IR} > 10^{12}L_{\odot}$).
Due to the small redshift range sampled by IRAS, these results can only be indicative,
though the suggestion that ULIGs might have played a stronger role in the past is 
supported by the detection of the strong infrared background. \\
With a thousand times better sensitivity and sixty times better resolution than IRAS, 
ISOCAM instrument (Cesarsky et al. 1996) on board of the {\it Infrared Space Observatory} 
(ISO; Kessler et al. 1996) has provided Deep and Ultra-Deep mid-infrared extragalactic 
surveys (mainly with the LW3 filter: 12 -- 18 $\mu$m), unveiling most of the star-formation 
in the Universe to $z = 1$. The source counts derived from these Deep/Ultra-Deep surveys 
(covering the flux density range 0.05 -- 4 mJy) strongly diverge from no-evolution models
at fluxes fainter than about 1 mJy, with an increasing difference that reaches a factor of 10 
around the faintest limits (0.05 -- 0.1 mJy; Elbaz et al. 1999a,b). The faint mid-infrared 
sources detected in the Deep/Ultra-Deep ISOCAM surveys have been identified mainly with
galaxies at $z \simeq 0.7$ and show LIG-like luminosities (Aussel et al. 1999b; Elbaz et 
al. 1999a,b; Flores et al. 1999). 

Although the Deep/Ultra-Deep ISOCAM Surveys have produced crucial results on galaxy evolution 
in the infrared, having identified most of the galaxies producing the mid-infrared background, 
there is a large gap in the flux density sampled by these surveys and 
by the IRAS Surveys. In particular, the flux density range 4 -- 200 mJy is essentially
uncovered in the Mid/Far--infrared and only a few sources have been detected by the
Deep ISOCAM Surveys in the important flux range 1--4 mJy, where most of the existing
evolutionary models (i.e. Franceschini et al. 2001; Xu 2000; Chary \& Elbaz 2001; 
Rowan-Robinson 2001) predict a substantial change
in the relative contribution of a local non-evolving and a more distant evolving population.
This scarcity of data in this flux interval reflects also in a poor knowledge of the local 
luminosity function for starburst galaxies and a corresponding uncertainty in the evolutionary 
properties of the different classes of infrared extragalactic objects.

The European Large Area ISO Survey (ELAIS; Oliver et al. 2000) is the largest survey 
conducted with the Infrared 
Space Observatory (ISO) and provides a link between the IRAS survey and the Deep/Ultra-Deep 
ISOCAM surveys. ELAIS is a collaboration between 20 European
institutes which involves a deep, wide-angle survey at high galactic
latitudes, at wavelengths of 6.7 $\mu$m (LW2), 15 $\mu$m (LW3), 90 $\mu$m (C100) and 
175 $\mu$m (C200) with ISO. In particular, the 15 $\mu$m survey covers a total area of 
$\sim$ 12 deg$^{2}$, divided into 4 main fields and several smaller areas. One of the 
main fields, S1, is located in the southern hemisphere. 
The whole S1 area has been surveyed in the radio (at 1.4 GHz, Gruppioni et al. 1999),
in several optical bands (La Franca et al. 2002, in preparation) and in the hard
X-ray with {\it BeppoSAX} (Alexander et al. 2001). Moreover, spectroscopic information and 
redshifts are available for a large number of sources (Gruppioni et al. 2001; La Franca, 
Gruppioni, Matute et al. 2002, in preparation).

We have reduced the 15$\mu$m data in S1 using a new ISOCAM data reduction technique 
({\it LARI technique}) especially developed for the detection of faint sources, obtaining
a catalogue (complete at the 5$\sigma$ level) of 462 sources in the flux density range 
0.45 -- 150 mJy. Details about the data reduction technique and the source catalogue have
been presented in Lari et al. 2001 (hereafter Paper I). 
Here we present the source counts at 15 $\mu$m derived from that catalogue.
The paper is structured as follows. In section \ref{survey} we give a brief description of
the 15 $\mu$m ELAIS survey in S1. In section \ref{compl} we present the completeness and
reliability of our sample at different flux levels. In section \ref{descr_cnt} we present 
the sample used to derive the source counts, which are shown in section \ref{counts}.
Finally, in sections \ref{discuss} and \ref{concl} we discuss our results
and their implications and present our conclusions. 

Throughout this paper we will assume $H_0 = 50$ km s$^{-1}$ Mpc$^{-1}$, $\Omega_m = 0.3$ and
$\Omega_{\Lambda} = 0.7$.

\section{Description of the ELAIS S1 Survey}
\label{survey}
The ELAIS survey at 15 $\mu$m, performed in raster mode with the ISOCAM instrument on board 
of ISO, covers a total area of $\sim$ 12 deg$^{2}$ divided into 4 main 
fields and several smaller areas. The main field located in the southern hemisphere (S1) is 
centered at $\alpha$(2000) = 00$^h$ 34$^m$ 44.4$^s$, $\delta$(2000) = $-43^{\circ}$ 28$^{\prime}$ 
12$^{\prime \prime}$ and covers an area of $2^{\circ} \times 2^{\circ}$. The 15 $\mu$m survey 
performed in S1 with the ISOCAM instrument consists of 9 different rasters.
Each raster covers an area of $\sim 43.5^{\prime} \times 42^{\prime}$; eight of them have been 
observed once, while one, S1$\_$5, was observed three times. 

The 15 $\mu$m data have been reduced and analyzed using the {\it LARI technique}, described
in detail in Paper I. With this data reduction method, we have obtained a sample of
462 sources with signal-to-noise ratio $\geq 5$   
in the flux range 0.45 -- 150 mJy. The fainter sources have been detected 
in the central raster of S1 (S1$\_5$), whose image has been obtained by combining 
three single observations centered on the same position.
The source catalogue in S1 and the relative parameter errors have been presented and discussed in 
Paper I, together with the detection rates at different flux densities derived with simulations.

The detection rates given in Paper I cannot be directly translated to completeness of the 
real catalogue, because our simulations were performed at discrete flux values rather
than following a continuous flux distribution. However, the results presented in Paper I
can be used to obtain the completeness of the catalogue and the source counts 
corrections, as discussed in the next section.

\section{Completeness and Reliability}
\label{compl}

\subsection{Brief Summary of the Results and Definitions from Paper I}
\label{summary}
Before describing in detail the method used to derive the completeness
of our source counts, it is useful to summarize the more relevant results 
and definitions of Paper I.
The simulations performed in the S1 field provided not only the completeness
of our detections at different flux levels, but also the internal calibration
of the source photometry and the distribution of the ratio between
the measured and the theoretical peak fluxes (crucial for the computations
described in next section).\\
Here we give some relevant definitions and relations:
\begin{itemize}
\item $f_s$: is the peak flux measured on maps for both real and simulated sources.
Its value depends both on data reduction method and on ELAIS observing strategy 
$+$ ISOCAM instrumental effects;
\item $f_0$: is the `theoretical' peak flux measured on simulated maps containing
neither glitches or noise. Its value depends only on ELAIS observing strategy
 $+$ ISOCAM instrumental effects;
\item $q \equiv f_s / f_0$;
\item $q_{med}$: is the peak of the $f_s / f_0$ distribution (also called
systematic flux bias) and is 0.78 $\pm$ 0.03 in S1 and 0.82 $\pm$ 0.03 in S1$\_5$.
These values are used to correct the measured flux densities;
\item flux density determination: 
\begin{equation}
s = \left(\frac{f_s}{f_0} \times s_0\right) / q_{med} \label{ftot}
\end{equation}
where for simulations $s_0$ is the injected total flux, while for real data
$s_0$ is derived through successive iterations starting from a rough estimated
value: 
\begin{equation}
s_0 = f_s / \left<f_s/s_0\right>_{sim} \label{flux}
\end{equation}
where $\left<f_s/s_0\right>_{sim} = 0.132$ is the average value resulting from simulations.\\
We can consider $s$ as the measured flux density and $s_0$ as the `true' flux density
of a source;
\item the photometric accuracy of our data reduction of the S1 area has been tested
using the stars of the field and following the relation calibrated on IRAS data by 
Aussel \& Alexander (2001) to predict the fluxes of stars. By a comparison between
the predicted fluxes and the ones derived from our analysis, we have obtained a very 
good agreement and a relative flux scale of 1.096 $\pm$ 0.044 (i.e. our fluxes have to
be multiplied by 1.096 to be put on the same scale as IRAS fluxes).
\end{itemize}

\subsection{The {\it g} Function}
\label{gfunc}
As described in Paper I and mentioned in the previous section, with simulations in 
S1 we have derived the distribution of
measured ($f_s$) to theoretical ($f_0$) peak flux ratio. This distribution is crucial
in deriving the completeness of the catalogue and the internal flux calibration. 
To this purpose, we have considered the 
$f_s / f_0$ distribution as a model function, hereafter called $g~ function$.
The $g$ distribution function, obtained for simulated sources, is a combination of an
intrinsic $g~ function$ (referred as $g_0$) plus a term due to noise.
As a rough estimate of $g_0$ in Paper I we considered
the measured $g$ distribution obtained for 3 mJy simulated sources ($g_{0}\_{_3}$). 
Because 3 mJy is a relatively high flux density (it is the highest flux injected in 
our simulations), the $g_{0}\_{_3}$ distribution is almost unbiased by detection incompleteness,
and we can consider it to be a good approximation of the intrinsic one. However, 
using the same simulation, we have obtained a more refined estimate of the intrinsic
$g_0$ function correcting the observed $f_s / f_0$ distribution for the low level of
detection incompleteness still present at this flux. Our procedure was the following:\\
each detected source $i$ has a $q_i = f_s(i) / f_0(i)$ value.
On a different position $j$, assuming that the 
$f_s(i) / f_0(i)$ value is not depending on position, the peak flux would be 
measured with a value $f_{s,i}(j) = q_i \times f_0(j)$ and would be detected only if 
$f_{s,i}(j)$ were greater than 5 times the
local noise ($\sigma(j)$). If $N_{i}$ is the number of possible detections in all the 
different $j$ positions of the simulations, the weight to be assigned to the $i^{th}$ 
source is $1/N_i$. Finally, the estimate of the intrinsic $g_0$ function
in a given interval of $q$ is obtained by summing the weights of all the detected sources
with a $q$ value in the same interval. By construction, the integral of $g_0$ over the 
entire range of $q$ is one.

\nin Once obtained the $g_0$ distribution, the general $g$ distribution can be derived
by convolving $g_0$ with the distribution of noise, here assumed to be Gaussian. This 
is possible in the approximation that the measured peak flux $f_s$ is the sum of a 
``true'' value, $f_s^t$, and a stochastic term, $\chi$, due to noise, so that the 
$q$ ($\equiv$ $f_s$/$f_{0}$) measured value is:

\begin{equation}
q = \frac{f_{s}}{f_{0}}=\frac{f_{s}^{t}}{f_{0}}+\frac{\chi}{f_{0}}
\end{equation}

\nin with the $\chi$ term following a Gaussian distribution with average equal to zero and
dispersion equal to the rms noise value, $\sigma$.
Following this formalism, then the $g$ distribution, expressed as a function of $f_{s}/f_{0}$ 
and $f_{0}/\sigma$, is obtained by just convolving $g_0$ with
the noise distribution, integrating over the possible values of the variable $\chi$:

\begin{eqnarray}
\label{g_distribution_eq}
g(\frac{f_{s}}{f_{0}},\frac{f_{0}}{\sigma})=\sqrt{\frac{1}{2 \pi}}\frac{1}{\sigma}\int{g_0(\frac{f_{s}}{f_{0}}-\frac{\chi}{
f_{0}})} \cdot e^{-\frac{\chi^2}{2 \sigma^2}}d\chi %= \\\nonumber
%\sqrt{\frac{1}{2 \pi}}\frac{1}{\sigma}\int{g_0(\frac{f_{s}}{f_{0}}-(\frac{\chi}{\sigma})\frac{1}{s_{0}})} \cdot e^{
%-(\frac{\chi^2}{2 \sigma^2})}d\chi=\\\nonumber
%\sqrt{\frac{1}{\pi}}s_{0}\int{g_0(\frac{f_{s}}{f_{0}}-\sqrt{2} x) \cdot e^{-x^2s_{0}^2}}dx \\\nonumber
%\frac{1}{\sqrt{\pi}} \frac{f_0}{\sigma} \int{g_0(\frac{f_{s}}{f_{0}}-\sqrt{2} x) \cdot e^{-x^2 (\frac{f_0}{\sigma})^2}}dx \\\nonumber
\end{eqnarray}

%\nin with $x=\frac{\chi}{\sqrt{2} f_{0}}$. 

\begin{figure}
\centerline{\psfig{figure=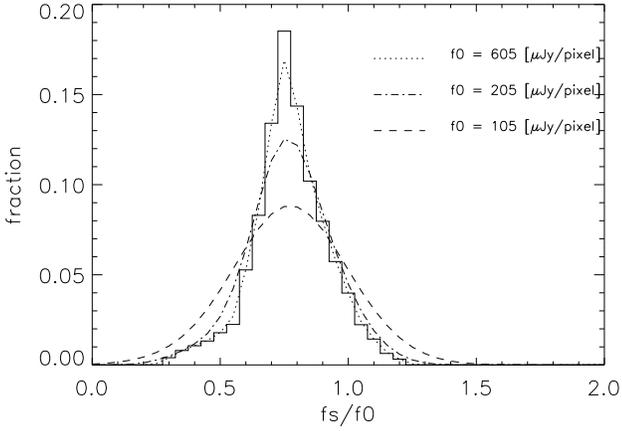,width=9cm}}
\caption{Predicted distributions of the ratio between the measured
and the theoretical peak flux, $f_s/f_0$ (also $g~ function$), 
in presence of a rms noise of 26 $\mu$Jy/pixel. These distributions, 
obtained from simulations, are shown for three different average values
of $f_0$ corresponding to different input fluxes. The solid histogram shows 
the intrinsic $g_0$ distribution (see text)}
\label{fsf0}
\end{figure}

\nin In figure \ref{fsf0} we show the intrinsic $g_0$
distribution computed as above (solid histogram) and
the predicted distributions of the ratio $f_s/f_0$
in presence of a $\sigma$ of 26 $\mu$Jy/pixel (typical of our data) for three different values 
of $f_0$, corresponding to different mean values of $f_0$ for different total input fluxes. 
%(i.e. $<f_0> = 605~ \mu$Jy/pixel for 3 mJy input flux). 
As we can notice, the presence of noise broadens the flux distributions 
and this effect becomes stronger towards fainter fluxes (as shown also in figure 7 of Paper I). 

\nin If we assume that the {\it g function} reflects all the multiplicative and additive 
error components due to the data reduction, we can use this function to predict the distribution 
of the detection rate for both simulated and real sources, as described in the next subsection.

\subsection{Completeness}

First we have computed the incompleteness introduced by our data reduction method, 
represented by the loss of sources not accounted for by our model. In fact, sources can be 
missed by our method if interpreted as background transients. The incompleteness of our method is
obtained from simulations by computing the ratio between the number of detections and the number
of expected sources in different peak flux intervals ($[f_{0}-df_{0},f_{0}+df_{0}]$).
The number of expected sources is derived by summing together the predicted detection
contributions ($I_{i}(f_{0}) = \int_{5 \sigma/f_{0}}^{\infty}{g(q,f_{0})dq}$) of all the sources 
$i$ with peak flux $f_{0i}$ belonging to the same interval:
\begin{equation}
N_{exp}(f_{0})=\sum_{f_{0i}\in[f_{0}{\pm}df_{0}]}{I_{i}(f_{0})}
\end{equation}
%
%dividing the number of detected sources by the number of predicted
%sources in different peak flux intervals [$f_0 - df_0$, $f_0 + df_0$]. The number of expected
%sources is obtained by summing together the contribution of all the $i$ detected sources with 
%$f_{0i}$ belonging to the interval $[f_{0}-df_{0},f_{0}+df_{0}]$:
%
%\begin{equation}
%N_{exp}(f_{0})=\sum_{f_{0i}\in[f_{0}{\pm}df_{0}]}{I_{i}(s_{0})}
%\end{equation}
%
%\nin where $I_{i}(s_{0}) = \int_{q_{min}(s_{0})}^{\infty}{g(q,s_{0})dq}$ is the prediction
%corresponding to the $i$ source with $f_{0}/\sigma = s_{0}$ and $q_{min}(s_{0})=5\sigma/f_0$ 
%is the minimum value of $q$ that corresponds to a detection, i.e. $f_s=5\sigma$.\\
\begin{figure}
\centerline{\psfig{figure=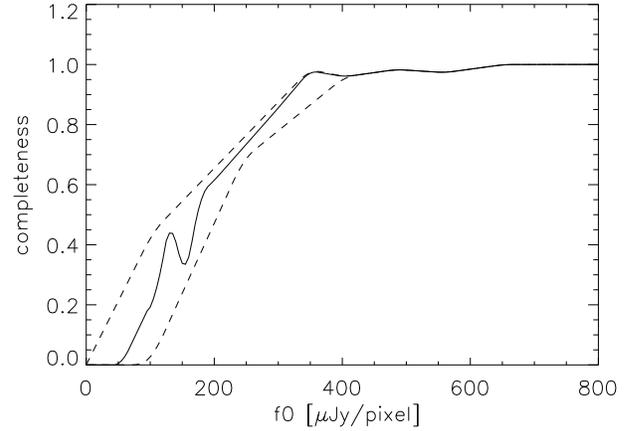,width=9cm}}
\centerline{\psfig{figure=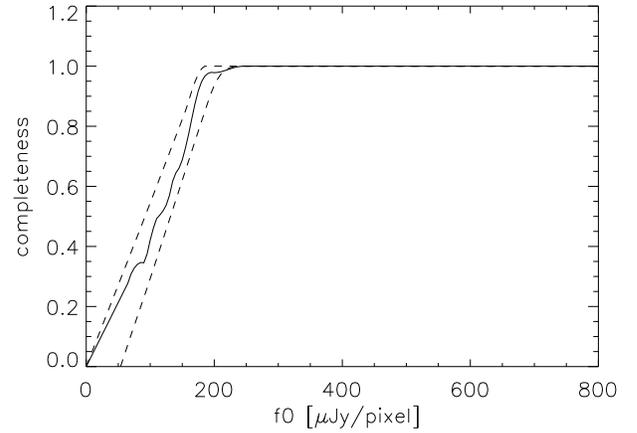,width=9cm}}

\caption{Completeness function of {\it Lari method} as
function of the theoretical peak flux for S1 ($top$) and S1$\_5$ ($bottom$).
The dashed lines give an estimate of the lower and upper envelopes of the completeness function.}
\label{co_fig}
\end{figure}

In figure \ref{co_fig} the resulting function describing the incompleteness of our method 
is plotted as a function of $f_{0}$, together with its lower and 
upper envelopes, for S1 ($top$ panel) and S1$\_5$ ($bottom$ panel).

To obtain the global correction to be applied to our source 
counts, we need to consider also the areal coverage of our survey (i.e. the fraction of the survey 
area where a source of peak flux $f_s$ can be detected: $f_s\ge5\sigma$) and the fact that real 
sources do not follow a discrete flux distribution as our simulated sources. To account for the 
latter effect, we have assumed a certain shape to describe the real source counts observed in the 
sky. According to the published counts at 15$\mu$m (e.g. Elbaz et al. 1999$a$), we have 
assumed two power laws between 0.4 and 150 mJy:

\begin{equation}
\label{dnds_ab}
\frac{d{\cal N}}{ds}(s) \propto \left\{ \begin{array}{ll}
                   S^{-\alpha_1} & \mbox{if $S>2$ mJy} \\
                   S^{-\alpha_2} & \mbox{if $S<2$ mJy}
                   \end{array}
             \right.  
\end{equation}

\nin with $\alpha_1$=2.3 and $\alpha_2$=3.0 as first estimate values.\\
By weighting the above 'theoretical' counts per unit of area by the $g~ function$ convolved with
the areal coverage function, and with the function describing the completeness of our method, 
we have computed the counts predicted in our Survey.
Being $dN(s) = dN(s)/ds \cdot ds$ the number of sources detected in the flux density bin
$ds$, $d{\cal N}(s_0) = d{\cal N}(s_0)/ds_0 \cdot ds_0$ the theoretical number
of sources detected in the `true' flux density bin $ds_0$ and $ds_0/ds = q_{med}/q$
(from equation \ref{ftot}), we have:
\begin{eqnarray}
\label{dnds_0_eq}
\frac{dN}{ds}(s)=\left<\int_{0}^{\infty} dq~ \left[g_{A}(q,f_0(s,q)) \cdot C(f_0(s,q)) \cdot \frac{d{\cal N}}{ds_0}(s_0) \cdot \frac{ds_0}{ds}\right] \right> \\
= \left<\int_{0}^{\infty} dq~ \left[g_{A}(q,f_0(s,q)) \cdot C(f_0(s,q)) \cdot \frac{d{\cal N}}{ds_0}(s_0)\cdot \frac{q_{med}}{q}\right] \right> \nonumber \\\nonumber
\end{eqnarray}

\begin{figure}
\centerline{\psfig{figure=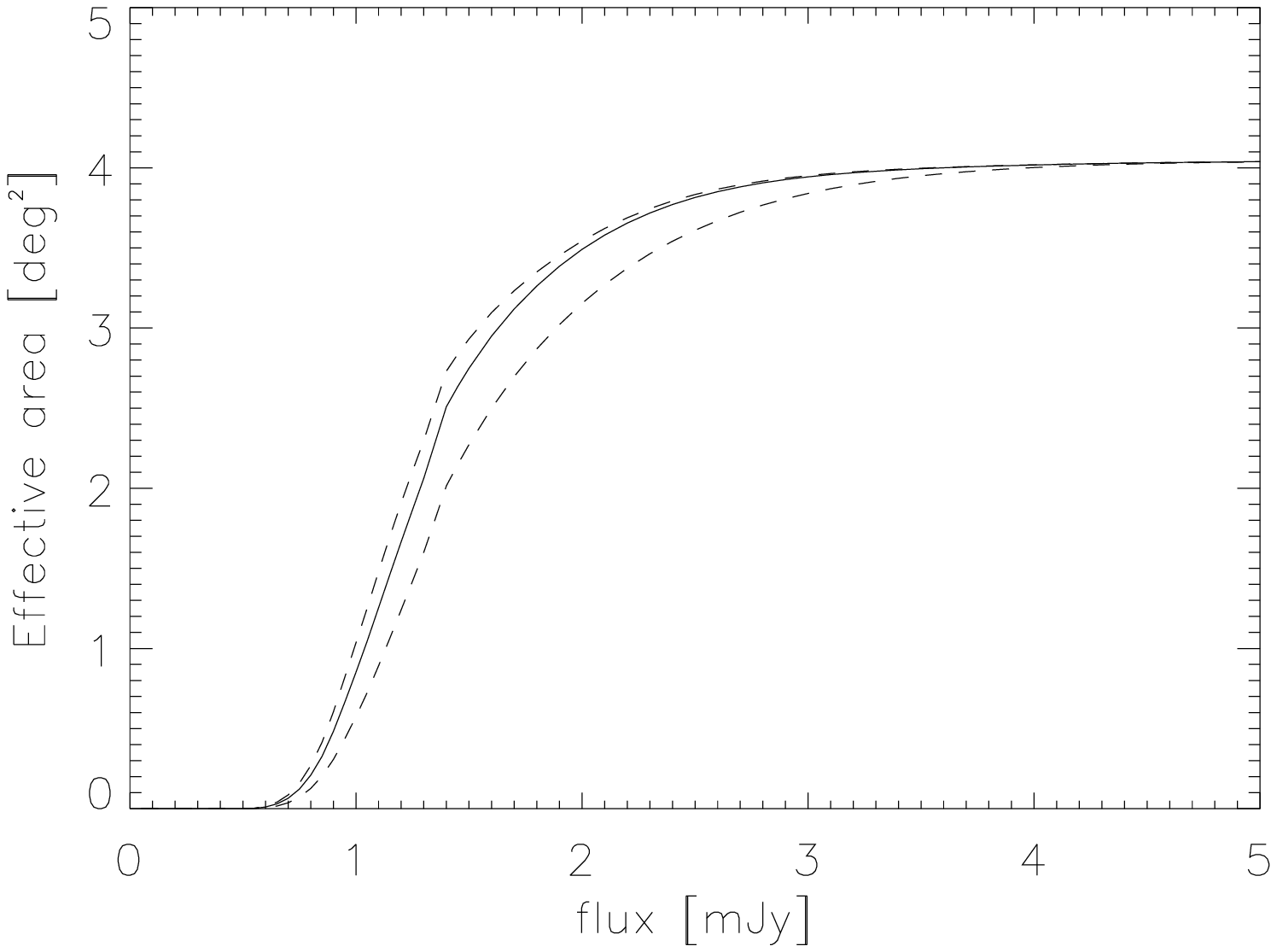,width=9cm}}
\centerline{\psfig{figure=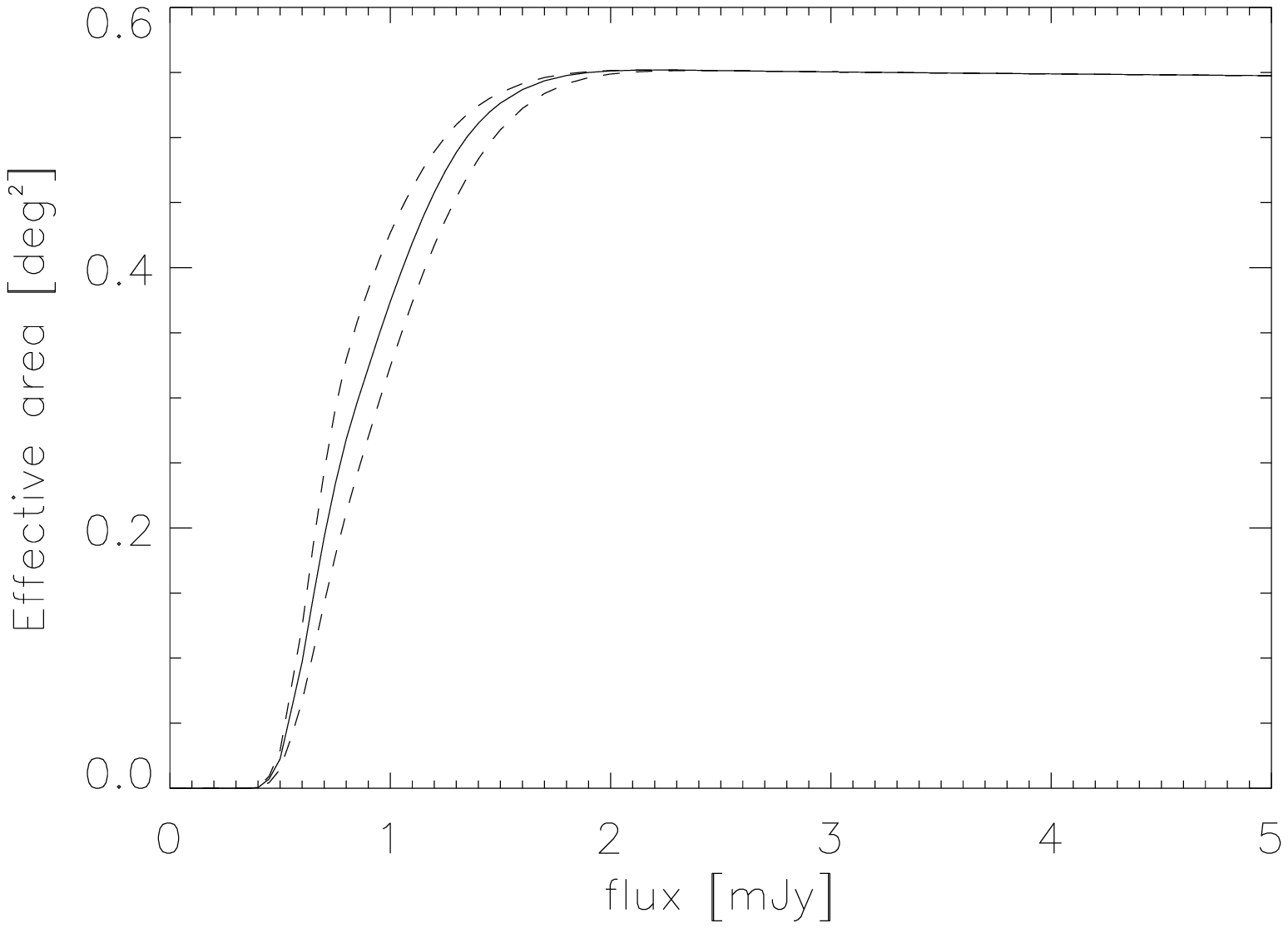,width=9cm}}

\caption{Effective area versus total flux density for the 15
$\mu$m catalogues in S1 main area ($top$) and S1$\_5$ ($bottom$).}
 \label{fig_compl}
\end{figure}

\normalsize
\nin where 
%$q_{med}$ (= 0.78 for S1 and 0.82 for S1$\_5$) is the systematic flux bias derived with
%simulations ($f_s$ is systematically lower than $f_0$ because of a loss of flux caused by data 
%reduction with {\it Lari method}; see Paper I), 
$g_A$ is the convolution of the $g~ function$ with the areal coverage function 
and $C$ is the completeness function shown in figure \ref{co_fig}. 
The averaging is performed over all the predicted 
peak flux values $f_0$ and with a random sampling of source positions in the survey. 
These expected source counts thus take into account the effects produced by the specific 
observational parameters of the ELAIS 15 $\mu$m survey and by our data reduction method. 
By putting $d{\cal N}(s_0)/ds_0$ in the form given by equation \ref{dnds_ab} into equation 
\ref{dnds_0_eq}, and changing $s_0$ to $s \cdot q_{med}/q$ (from equation \ref{ftot}), 
we obtain the source counts predicted for our survey, $dN(s)/ds$. The ratio between these 
counts and the theoretical source counts, $d{\cal N}(s_0)/ds_0$, gives the global correction 
(including incompleteness and areal coverage) to be applied to our measured source 
counts. The inverse of the global correction ({\it effective area}) is reported in figure 
\ref{fig_compl} for both $S1$ ($top$ panel) and $S1\_5$ ($bottom$ panel) surveys. 
Since our source counts corrected for incompleteness were significantly different
from the original model counts (especially the power law at faint flux densities), we have 
iterated this procedure by adjusting the power laws parameters until convergence
is achieved. The final estimate of the model counts $d{\cal N}(s)/ds$ is obtained 
for $\alpha_1$=2.3 and $\alpha_2$=3.6.  

\section{The Data Sample Used for the Source Counts}
\label{descr_cnt}

The catalogue used to derive the source counts is not exactly the catalogue published 
in Paper I, but contains some differences, as described in this section.

First, we have conservatively decided to exclude from the source list used for counts 
35 sources detected in S1, all with flux density $< 1.5$ mJy, resulting ``dubious'' at 
a visual inspection of their pixel history. These sources, detected above the 5$\sigma$
threshold on the maps obtained through a combination of several images, are too faint
to be distinguished from noise on the single pixel histories without uncertainty. Moreover,
an additional factor which strengthened our doubts about the reliability of these 
sources is their very low optical identification rate. In fact, while for the entire
catalogue (minus the 35 ``dubious'' sources) the optical identification rate 
(within a circle of 4 arcsec radius) on the DSS2
images for sources fainter than 1.5 mJy is 60\%, for the 35 ``dubious'' sources it is only 14\%, 
with a chance detection rate of about 10\%.

\begin{figure}
\centerline{\psfig{figure=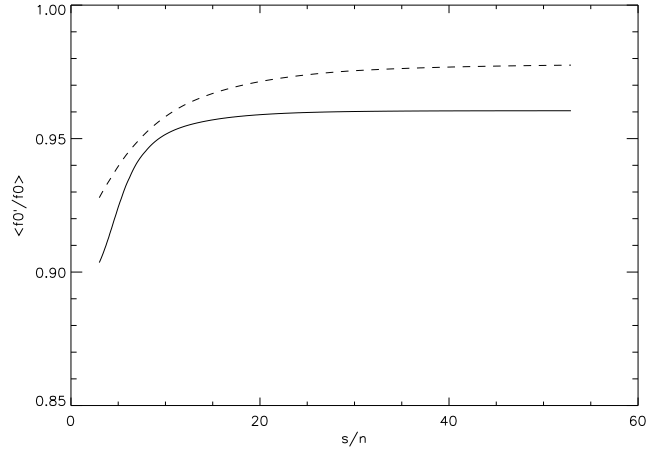,width=9cm}}
\caption{Correction factors as function of signal to noise to be applied to the
$S1$ ($solid$) and $S1\_5$ ($dashed$) source flux density to account for flux loss caused
by positional errors in the theoretical peak flux $f_0$.}
\label{f0_sn}
\end{figure}

Second, before computing the source counts we have applied three small
further corrections to the flux densities presented in the catalogue of Paper I:
\begin{itemize}
\item [1.] We have applied the calibration factor of 1.096 derived from the comparison 
with stars in order to put our fluxes on the same scale as IRAS fluxes (see section \ref{summary} 
and section 7 in Paper I).
\item [2.] We have corrected for the average underestimate of the true flux introduced by 
positional errors of the auto-simulated peak flux $f_0$ (see section 6 in Paper I); Since 
$f_0$ is computed on the measured positions and not on the `true' 
ones, the measured flux is on average underestimated. This underestimate is flux dependent, due to
the flux dependence of positional uncertainties. 
In figure \ref{f0_sn} the corrections to be applied to the source flux densities
due to this effect are reported as a function 
of signal to noise for both $S1$ (solid line) and $S1\_5$ (dashed line).
\item [3.] In $S1\_5$ only we have corrected for the additional loss of flux due to the 
combination of the three rasters.
To estimate this correction factor, for each source found in the combined $S1\_5$ map
we have measured the flux in the three separate rasters and compared their average with the 
flux measured on the combined map. The mean ratio between the combined and the averaged 
single fluxes, considered as the correction factor, is $R = 0.96 \pm 0.01$.  
\end{itemize}

In the following statistical analysis we have chosen not to eliminate the 20 repeated sources
(those belonging to the overlapping regions of two different rasters), but to consider
them as different sources. The reason of this choice is that the detectability and
completeness analysis, and consequently the correction factors to be applied to our data, 
have been performed on the single raster areas. This can explain why the effective area
of S1 (see figure \ref{fig_compl}) corresponding to bright fluxes (= 8 times the area
of one raster) is larger than the area of sky effectively covered by the S1 rasters.
The sum of the effective areas shown in figure \ref{fig_compl} is about 4.6 square degrees,
while the S1 $+$ S1$\_5$ survey covers an area of 4 square degrees (about 15\% of
the area is covered by at least two rasters).

Finally, to compute the extragalactic
source counts, we have excluded from our lists all the sources with a stellar counterpart brighter
than $J = 16$ in the Guide Star Catalogue II\footnote{The Guide Star Catalogue-II is a joint project 
of the Space Telescope Science Institute and the Osservatorio Astronomico di Torino. Space Telescope 
Science Institute is operated by the Association of Universities for Research in Astronomy, for the 
National Aeronautics and Space Administration under contract NAS5-26555. The participation of the 
Osservatorio Astronomico di Torino is supported by the Italian Council for Research in Astronomy. 
Additional support is provided by European Southern Observatory, Space Telescope European Coordinating 
Facility, the International GEMINI project and the European Space Agency Astrophysics Division.}
and with an evident stellar appearance (i.e. point-like with spikes) on the DSS2\footnote{The 
``Second epoch Survey'' (DSS2) of the Southern Sky was made by 
the AAO with the UK Schmidt Telescope. Plates from this survey have been digitized and compressed
by the STScI. The digitized images are copyright(c) 1993-1995 by the AAO Board and are distributed
herein by agreement. All rights reserved.} images.
We have chosen not to exclude any stellar identification fainter than $B_J = 16$ found in the GSC-II 
(which is complete to $B_J = 19.5$), because the reddest faint 
stars in our sample have estimated $B_J$ magnitude of the order of 15 (Aussel and Alexander 2001), 
and at fainter magnitudes the elimination from the sample of stellar-like objects might cause 
the elimination of AGN instead of stars. In fact, at $B_J \gsimeq 16$ the identifications 
with point-like objects are expected to be dominated by AGN. \\
\begin{figure}
\centerline{\psfig{figure=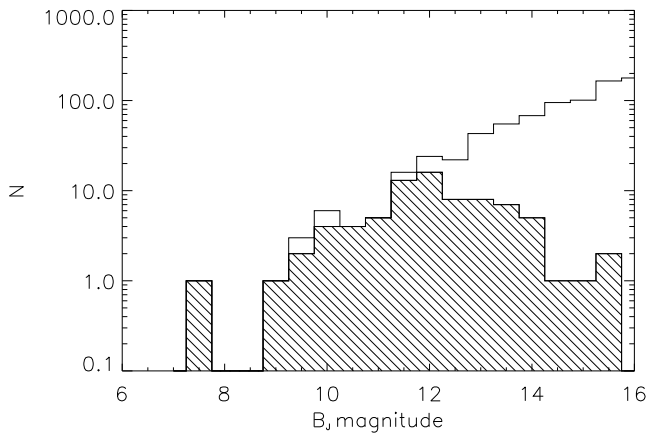,width=9cm}}
\caption{Magnitude distribution of our stellar identifications (line-filled
histogram) to $B_J = 16$, compared to the magnitude distribution of all the 
GSC-II stars in the S1 area (empty histogram).}
\label{star_mag}
\end{figure}
\begin{figure}
\centerline{\psfig{figure=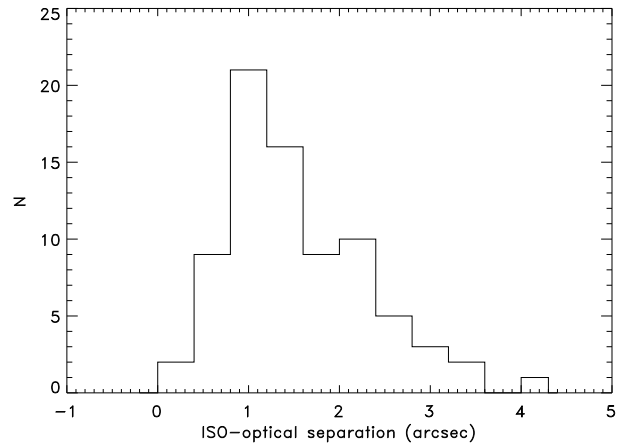,width=9cm}}
\caption{ISO-optical separation distribution for the stellar identifications
in S1.}
\label{star_sep}
\end{figure}
In the end, we have identified with stars, and subtracted from our list, 82 sources in S1 and 20 
in S1$\_5$ (in total 102 stars, 87 of which are different), thus leaving a total of 325 
extragalactic sources in S1+S1$\_5$ (320 different). The $B_J$ magnitude 
(from GSC-II) distribution of these stars is reported in figure \ref{star_mag} (filled histogram), 
where also the magnitude distribution of all the GSC-II stars in the S1 area is reported (to $B_J 
= 16$). The maximum separation we find between ISO sources and star positions for our stellar
identifications is $\lsimeq 4$ arcsec, as shown in figure \ref{star_sep}.
Given these positional differences and the surface density of stars, we estimate that less 
than one ISO/star association 
could be spurious (down to the considered magnitude limit $B_j=16$). \\
In figure \ref{star_frac} the fraction of stars in S1 is plotted as a function of flux density.
At $S_{15~\mu m} \gsimeq 50$ mJy all our sources are identified with stars and still at $S_{15~\mu m} 
\simeq 5$ the fraction of stars is $\geq 50$\%. For this reason, an accurate star subtraction is
very important before computing the extragalactic ISOCAM counts, even at relatively faint fluxes
(1-2 mJy), where the fraction of stars is still of the order of 20\%.    

\begin{figure}
\centerline{\psfig{figure=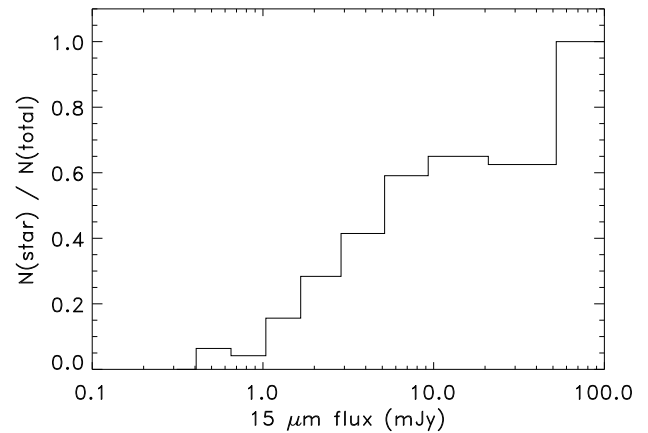,width=9cm}}
\caption{Ratio between stars and the total number of 15 $\mu$m sources
as a function of flux density for our S1 sample.}
\label{star_frac}
\end{figure}

In the end, the extragalactic sample used to compute the source counts is composed
by 325 source, 320 of which are different.
%This revised sample, with the above corrections
%applied to the flux densities, can be obtained from
%{\it http://boas5.bo.astro.it/$\sim$elais/catalogues/counts$\_$sample.html}.

\section{Source Counts}
\label{counts}

Given the extent of the 15 $\mu$m survey in S1 and the significant depth reached in its 
central area, 
S1$\_$5, our source sample is optimally suited to study the ISOCAM source counts with a large
statistics and over a broad flux range (0.5 - 100 mJy). Therefore, the combined sample of our 
S1(main area) + S1$\_$5 non-stellar sources with $S_{15\mu m} \geq 5\sigma$ has been used to
construct the mid-infrared extragalactic source counts distribution. 

We used the effective areas derived in section \ref{compl} 
(see figure \ref{fig_compl}), to obtain the
extragalactic mid-infrared source counts from our 15 $\mu$m samples. 
\tiny
\begin{table*}
\centering
\begin{minipage}{180mm}
  \caption{The 15 $\mu$m Source Counts in S1 + S1$\_5$}
  \label{counts_tab_tot}
\tiny
\begin{tabular}{ccrccrccrccc}
 \hline
& & & & &  \\
 & &\multicolumn{3}{c}{S1} & \multicolumn{3}{c}{S1$\_5$} &  \multicolumn{4}{c}{S1$+$S1$\_5$} \\
& & & & &  \\
$S$ & $\left< S \right>$ &$n$ & $A_{eff}$ & $dn/dS$ & $n$ &$A_{eff}$ & $dn/dS$ & 
 $n$ & $dn/dS$ & $dn/dS \cdot S^{2.5}$ & $N(>S)$ \\
& & & & &  \\
 (mJy) & (mJy) & & (deg$^{2}$) & (deg$^{-2}$ mJy$^{-1}$) & & (deg$^{2}$) & (deg$^{-2}$ mJy$^{-1}$) &
 &  (deg$^{-2}$ mJy$^{-1}$) & (deg$^{-2}$ mJy$^{1.5}$) & (deg$^{-2}$) \\
& & & & &  \\ \hline
  0.50 --  0.80  & 0.63 & 6 & 0.02 &           $-$   & 19 & 0.13 &1344 $\pm$ 539~  & 19 & 1344 $\pm$ 539~     &428 $\pm$ 171 &~544 $\pm$ 163\\  
  0.63 --  1.01  & 0.80 & 34& 0.21 & $672 \pm 157$ & 27 & 0.27 & 521 $\pm$ 113 & 61 &  645 $\pm$ 109      & 369 $\pm$ ~62 & ~353 $\pm$ ~42~\\
  0.80 --  1.28  & 1.01 & 76& 0.90 & $380 \pm 59~$  & 21 & 0.38 &~161 $\pm$ ~36~ & 97 &  280 $\pm$ ~33      & 288 $\pm$ ~34 & ~194 $\pm$ ~17~\\ 
  1.01 --  1.62  & 1.28 &88& 2.00 &  $127 \pm 5~~$  & 19 & 0.48 &~~81 $\pm$ ~~19~& 107 &  116 $\pm$ ~12     & 215 $\pm$ ~22 & ~108 $\pm$ ~~8~~\\
  1.28 --  2.05  & 1.62 &77& 2.94 &  $44~ \pm 5~~$  & 19 & 0.54 &~~51 $\pm$ ~~12~&  96 &   45 $\pm$   ~5    & 152 $\pm$ ~16 &  59.6 $\pm$ 4.4~\\
  1.62 --  2.75  & 2.11 &67& 3.57 &  $18~ \pm 2~~$  & 16 & 0.54 &~~26 $\pm$ ~~7~ &  83 &   20 $\pm$   ~2    & 126 $\pm$ ~14 &   37.8 $\pm$ 3.1~\\
  2.05 --  3.69  & 2.75 &50& 3.85 &  $8.1 \pm 1.2$ & 10 & 0.54 &~~11 $\pm$ ~~4~ &  60 &  8.6 $\pm$ 1.1     & 108 $\pm$ ~14 &   24.7 $\pm$ 2.4~\\
  2.75 --  4.95  & 3.69 &34& 4.00 &  $3.9 \pm 0.7$ &  4 & 0.54 & ~3.3 $\pm$  1.7 &  38 &  3.9 $\pm$   0.6   & 101 $\pm$ ~16 &  15.8 $\pm$ 1.9~\\
  3.69 --  6.64  & 4.95 &24& 4.00 &  $2.0 \pm 0.4$ &  2 & 0.54 & ~1.2 $\pm$  0.9 &  26 &  1.9 $\pm$   0.4   & 105 $\pm$ ~21 &  10.6  $\pm$ 1.5~\\
  4.95 --  8.90  & 6.64 &18& 4.00 &  $1.1 \pm 0.3$ &  1 & 0.54 & ~0.5 $\pm$  0.5 &  19 &  1.1 $\pm$   0.2   & 120 $\pm$ ~28 &  ~7.3  $\pm$ 1.3~\\
  6.64 --  11.9  & 8.90 &11& 4.00 &  $0.5 \pm 0.2$ &  0 & 0.54 & ~0.0 $\pm$  0.0 &  11 &  0.5 $\pm$   0.1   & 109 $\pm$ ~33 &  ~4.9  $\pm$ 1.0~\\
  8.90 --  18.9  & 13.0 & 7& 4.00 &  $0.2 \pm 0.1$ &  1 & 0.54 & ~0.2 $\pm$  0.2 &   8 &  0.2 $\pm$   0.1   & 106 $\pm$ ~38 &  ~3.1  $\pm$ 0.8~\\
  11.9 --  29.9  & 18.9 & 5& 4.00 & $0.07 \pm 0.03$ &  2 & 0.54 & ~0.2 $\pm$  0.1 &   7 &  0.09 $\pm$   0.03 & 133 $\pm$ ~50 &  ~2.2  $\pm$ 0.7~\\
  18.9 --  47.2  & 29.9 & 2& 4.00 &  $0.02 \pm 0.01$ &  1 & 0.54 & ~0.1 $\pm$  0.1 &   3 &  0.02 $\pm$   0.01 & 113 $\pm$ ~65 &  ~1.1  $\pm$ 0.5~\\
  29.9 --  74.6  & 47.2 & 3& 4.00 &  $0.02 \pm 0.01$ &  0 & 0.54 & ~0.0 $\pm$  0.0 &   3 &  0.02 $\pm$   0.01 &~225 $\pm$ 130 &  ~0.7  $\pm$ 0.4~\\
 ~47.2 -- 118.0  & 74.6 & 2& 4.00 & $0.007 \pm 0.005$&  0 & 0.54 & ~0.0 $\pm$  0.0 &   2 & 0.006 $\pm$  0.004 &~298 $\pm$ 211 &  ~0.4  $\pm$ 0.3~\\
 \hline      
\end{tabular}
\end{minipage}
\end{table*}
\normalsize
In Table~\ref{counts_tab_tot} the
15 $\mu$m source counts in S1, S1$\_5$ and S1+S1$\_5$ areas are presented. 
We have first computed
the counts for the two samples in S1 and S1$\_5$ separately, as reported in the first
eight columns of Table \ref{counts_tab_tot}, giving respectively 
the adopted flux density intervals, the average flux density
in each interval (computed as the geometric mean of the two flux density limits), the
observed number of sources, the effective area (figure \ref{fig_compl}$top$), the differential source 
counts and their associated errors in S1 and the observed number of sources, the effective area 
(figure \ref{fig_compl} $bottom$) and the differential source counts 
with their errors in S1$\_5$. The differential source counts have been obtained by weighting
each single source for its effective area rather than weighting the total number of sources in each flux 
density bin for the effective area corresponding to the reference flux density of that bin. 
The errors associated to the counts in each bin have been computed as $\sqrt{\Sigma_i (1/A_{eff}^2(S_i))}$, 
where the sum is for all the sources with flux density $S_i$ belonging
to the bin and $A_{eff}(S_i)$ is the effective area corresponding to that source flux.
These errors take into account only the Poissonian term of the uncertainties associated to
the source counts, for consistency with the other literature works. Especially at faint flux 
density, where the effective area (and consequently the correction factor) is a steep function
of flux, the errors quoted in Table \ref{counts_tab_tot} should be considered as lower limits
of the `true' errors (including also the uncertainty in the effective area computation,
shown as dashed curves in figure \ref{fig_compl}). \\
The counts computed for S1 and S1$\_5$ are consistent
within the errors in most flux density intervals, the only exception being the 0.8 - 1.2 mJy flux bin,
where the counts in S1 are larger than the ones in S1$\_5$ at a formal level of $\sim 2.3 \sigma$.
However, since the low flux density errors are somewhat underestimated, we can consider
the source counts in S1 and S1$\_5$ consistent in all the common flux density intervals, 
over the whole range 0.5 - 100 mJy. \\
For this reason, we have also computed the source counts in the ``combined'' sample (S1$+$S1$\_5$),
by considering all the sources as belonging to a unique sample: for each source the 
combined effective area is the sum of the effective areas in S1 and S1$\_5$ (whose values 
are reported in column 4 and 7 respectively). In the last four columns the total number of sources
in each flux density bin, the differential source counts, the differential counts normalized to the 
Euclidean distribution (by multiplying by $\left<s\right>^{2.5}$) with their errors and the integral 
source counts (with errors) for the combined sample are reported.   
In the first flux density bin the counts for the combined sample coincide with the $S1\_5$ 
counts, since, due to their negligible effective area, we have not considered the $S1$ data.

Note that the flux bins are partially overlapping, therefore they are not 
statistically independent (they are alternately independent).
The choice of partially overlapping flux density bins for our source counts
representation is based on the need of a tight sampling of the region where
the counts start diverging from no evolution models, in order to better determine
the break point and the counts shape (with a larger statistics).
As mentioned in the previous section, in computing the counts we have considered as two 
different sources those appearing
in two different rasters (belonging to the border part of a raster, overlapping with 
an adjacent raster), by suitably weighting them for their detectability
area in each raster. In fact, in deriving the areal coverage function we have considered
the total area of each single raster, including the overlapping regions. 

\begin{figure*}
\begin{minipage}{180mm}
\centerline{\psfig{figure=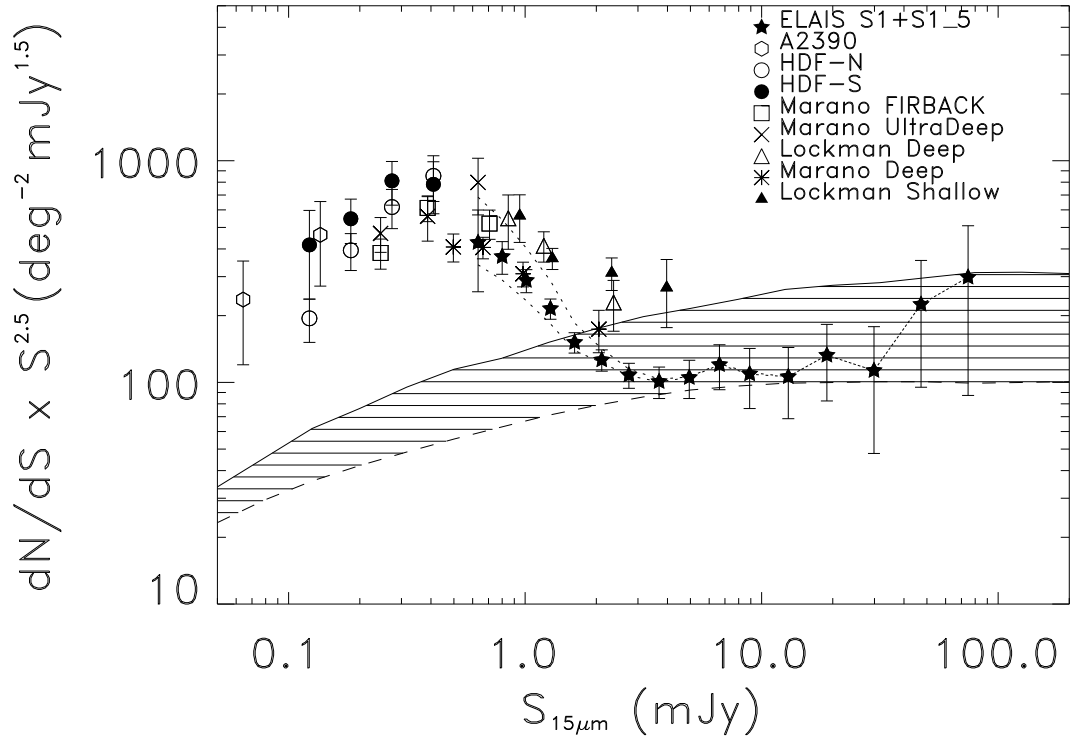,width=16cm}}
\caption{Differential source counts at 15 $\mu$m normalized to the Euclidean distribution 
($N(S) \propto S^{-2.5}$).
Data points (as shown also in the figure): A2390 (open diamonds), ISO HDF--N (open circles), ISO HDF--S
(filled circles), Marano Firback (open squares), Marano Ultra-Deep (diagonal crosses), Marano Deep 
(asterisks), Lockman Deep (open triangles), Lockman Shallow (filled triangles), ELAIS S1 (filled stars). 
The dotted curves are the lower and upper envelopes of our counts due to the uncertainties
in the completeness curve derivation, as shown in figure \ref{fig_compl}. The hatched area represents
the range of possible expectations from no evolution models normalized to the IRAS 12 $\mu$m Local 
Luminosity Function (upper limit from Rush, Malkan \& Spinoglio 1993; lower limit from Fang et al. 1998).}

\label{fig_diffcount}
\end{minipage}
\end{figure*}

The 15 $\mu$m differential source counts of the combined ELAIS S1 and S1$\_5$ 
data, normalized to those expected in a Euclidean geometry by dividing 
by $S^{-2.5}$, are shown in figure \ref{fig_diffcount} (filled stars).  
For comparison, source counts from other ISOCAM surveys (A2390 from Altieri et al. 1999; 
ISO HDF-N from Aussel et al. 1999a; ISO HDF-S, Marano Firback, Marano Ultra-Deep, 
Marano Deep, Lockman Deep, Lockman Shallow from Elbaz et al. 1999b; data kindly 
provided by D. Fadda, private communication) are also plotted. 
Our counts are lower
than the Lockman Deep and Shallow ones. However, they appear consistent with
the counts obtained in the Marano Deep Survey, at least in the common flux density range
(0.5 -- 2 mJy). 

\begin{figure*}
\begin{minipage}{180mm}
\centerline{\psfig{figure=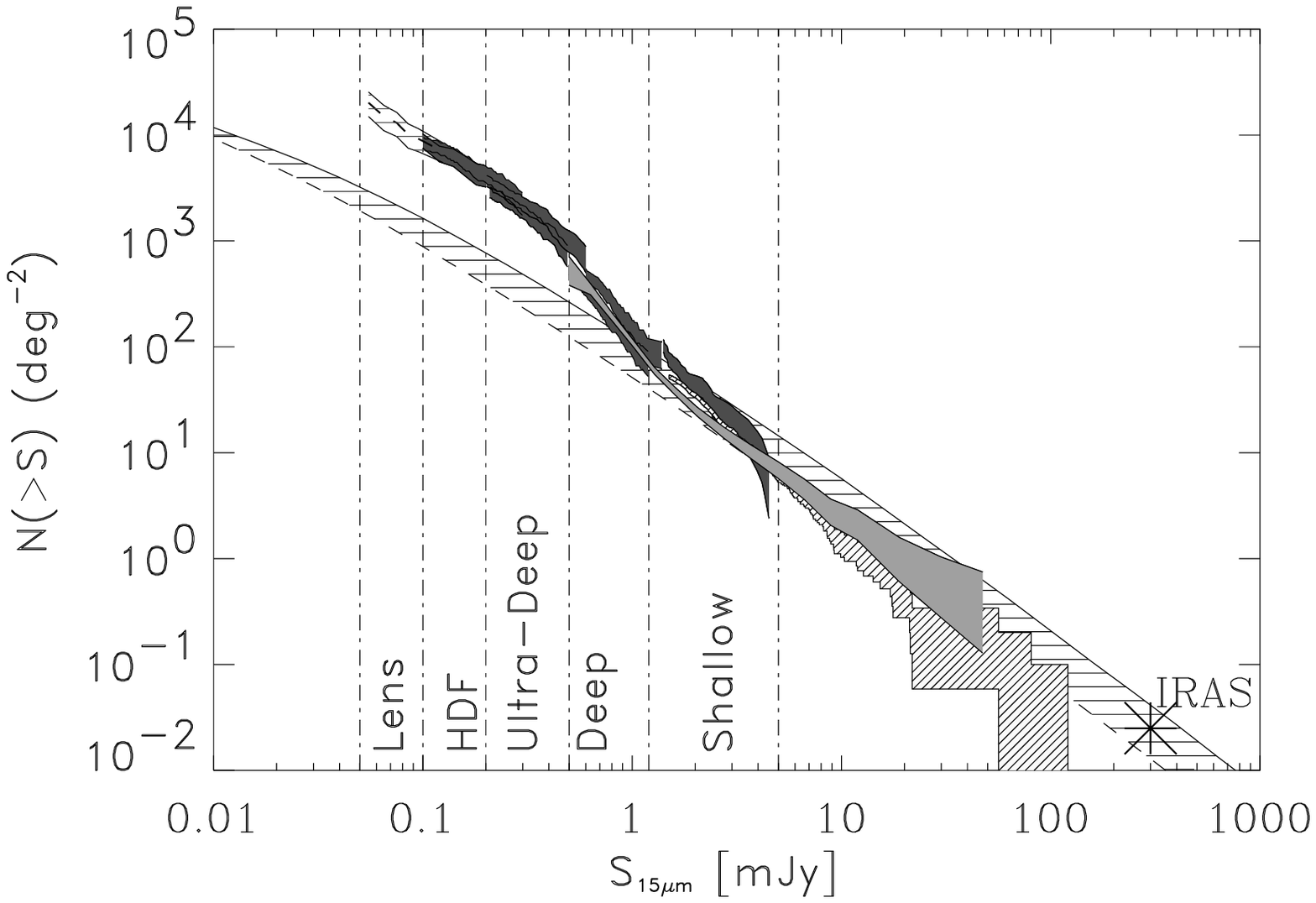,width=16cm}}
%\centerline{\psfig{figure=int_count_comb_rush_pa2.ps,width=18cm}}
\caption{Integral source counts for extragalactic ISOCAM sources detected at 15
$\mu$m above a given flux $S$.
The grey-shaded area shows the integral extragalactic counts with 68\% confidence contours
obtained from the ELAIS S1 Survey (this work). For comparison, the hatched 
and the black-shaded areas represent the counts
with 68\% confidence contours obtained respectively from the ELAIS Preliminary
Analysis, normalizing downwards by a factor of 2 the fluxes of Serjeant et al. (2000), 
and from the ISOCAM Deep/Ultra-Deep Surveys (Elbaz et al.
1999b). The area filled with horizontal lines represents the range of possible expectations 
from no-evolution models normalized to the IRAS 12$\mu$m local luminosity function, 
as reported in figure \ref{fig_diffcount}. The fainter end of the IRAS 12$\mu$m source counts 
derived by Rush, Malkan and Spinoglio (1993), converted to 15 $\mu$m, 
%are represented by the hatched area at $>$300 mJy.
is represented by the asterisk. 
%Apart from the overplotted ELAIS S1 and PA counts, this 
%figure is taken from figure 1 of Elbaz et al. (1999b).
}\label{fig_intcount}
\end{minipage}
\end{figure*}

In figure \ref{fig_intcount} our integral extragalactic source counts are reported.
The grey-shaded area shows the integral extragalactic counts with 68\% confidence contours 
obtained from the ELAIS S1 + S1$\_5$ Survey. For comparison, the black-shaded
area represents the counts with 68\% confidence contours obtained from the ISOCAM 
Deep/Ultra-Deep Surveys (Elbaz et al. 1999b). The hatched area represents the integral 
counts of Serjeant et al. (2000) based on the Preliminary Analysis of the whole ELAIS 
Survey, with fluxes rescaled downward by an average factor of 2, as suggested by a recent 
calibration work (i.e. V\"ais\"anen et al. 2002). The area filled with horizontal lines 
represents the range of possible expectations from no-evolution models normalized to the 
IRAS 12$\mu$m local luminosity function. 
The asterisk represents the fainter end of the IRAS 12$\mu$m source counts derived 
by Rush, Malkan and Spinoglio (1993), opportunely converted to 15 $\mu$m. 
%are represented by the hatched area at $>$300 mJy. 

Our data reduction allowed us to compute source counts down to fluxes
$\sim$3 times fainter than the Preliminary Analysis (PA) of the ELAIS data
(Serjeant et al. 2000). In the flux range common to the two samples we find
that, after correcting downward the PA fluxes by a factor of 2, our source
counts are in reasonably good agreement with the counts of Serjeant et al. 
(2000) above $\sim$2.5 mJy. However, the overall slope of the
Serjeant et al. counts appears to be steeper than ours, with our counts
being lower below 2.5 mJy and somewhat higher above 10 mJy. This is
probably due the fact that a flux dependent calibration correction, rather
than a constant correction, should be applied to the Serjeant et al. (2000)
fluxes. In fact, there are hints (Babbedge and Rowan-Robinson 2002, in
preparation) that the needed correction factor is $\sim$2.4 for the fainter
PA fluxes and $\sim$1.75 for the brighter fluxes (the factor of 2 here adopted
is an average value).
 
Our counts, though not deep enough to detect the fast convergence at flux densities 
fainter than 0.4 mJy shown by the Deep/Ultra-Deep ISOCAM counts, sample very well the flux density 
range where those counts start diverging from no evolution models. Indeed, we observe a remarkable
change in the slope of our counts, showing a significant super-Euclidean slope from about 2
mJy to $\sim$0.45 mJy. Due to its large statistics, our Survey is at the moment best suited for 
determining both the exact flux density where the 15 $\mu$m extragalactic counts steepen and 
the count slope itself (before and after the steepening).
A maximum likelihood fit to our extragalactic source counts with
two power laws:
\begin{equation} 
 \frac{dN}{dS} \propto \left\{ \begin{array}{ll}
                   S^{-\alpha_1} & \mbox{if $S>S_b$} \\
                   S^{-\alpha_2} & \mbox{if $S<S_b$}
                   \end{array}
             \right.  
\end{equation} 
\nin gives the following parameters: $\alpha_1 = 2.35 \pm 0.05$, $\alpha_2 =
3.60 \pm 0.05$, $S_b =$($2.15 \pm 0.05$) mJy. 
Our best fit parameters suggest that the steepening of the integral counts 
starts around 2 mJy, then the counts keep a super-Euclidean slope down to
the limits of our survey ($\sim$0.5 mJy).

\section{Discussion}
\label{discuss}

\subsection{Comparison with Deep/Ultra-Deep ISOCAM Surveys Source Counts}
\label{deep}
The 15 $\mu$m extragalactic source counts derived from the southern ELAIS Survey 
cover over two decades in flux, from 0.5 up to 100 mJy, with a significant statistical 
sampling (325 objects). Due to the large flux density interval covered, the ELAIS counts
bridge the gap existing between the IRAS counts and the ISOCAM Deep/Ultra-Deep counts.
The ELAIS Survey was planned to be a shallow survey and to reach an optimistic 
limit of about 2 mJy. As shown in Paper I, with the {\it Lari technique} we were able to
go much deeper than expected, detecting a significant number of sources even at sub-mJy 
levels. 
%However, at fluxes fainter than $\sim$1 mJy the S1 data require a large incompleteness
%correction (a factor of $\gsimeq 4$), while the S1$\_5$ data, though more complete due to the
%repeated observations, have a much smaller statistical significance due to the smaller 
%area covered. The counts presented
%here are the weighted average of the S1 and S1$\_5$ source counts. We must note that 
The strength
of ELAIS counts is at fluxes brighter than $\sim$1 mJy, where they are highly statistically 
significant and complete. At fainter fluxes, though the S1 and S1$\_5$ counts are
consistent over the whole flux range, the results are less strong 
due to the large incompleteness correction required by the S1 data and the small area
covered by the more complete S1$\_5$ data. However, the results 
are consistent with the evolution scenario found by other ISOCAM surveys and are able to give 
general hints on the behaviour and evolution of infrared galaxies. In particular,     
ELAIS counts in the flux density range in common with ISOCAM Deep counts 
(0.5 -- 4 mJy)
diverge from no evolution models as well and steepen with a super-Euclidean slope (3.60 $\pm$ 
0.05 for the differential form) up to the fainter limit. In particular, the flux density 
where our counts start diverging from no evolution predictions is (2.15 $\pm$ 0.05) mJy. 
Above this flux density, the ELAIS counts are consistent with no evolution, showing a slope 
(in differential form) of 2.35 $\pm$ 0.05.
Although similar results have been found for the Deep ISOCAM Surveys,
between 0.5 and $\sim$2 mJy our counts are somewhat steeper and on average lower than the others. 
However, the common flux range is sampled by three Deep Surveys only:
Marano Deep, Lockman Deep and Lockman Shallow. The counts drawn from the Marano Deep Survey
appear consistent with the ELAIS counts, while the counts obtained from both Lockman 
Deep and Shallow Surveys are less steep at faint fluxes (especially the Lockman Shallow ones) 
and higher than ours by about a factor of 2-3 around 2-3 mJy.

\begin{figure}
\centerline{\psfig{figure=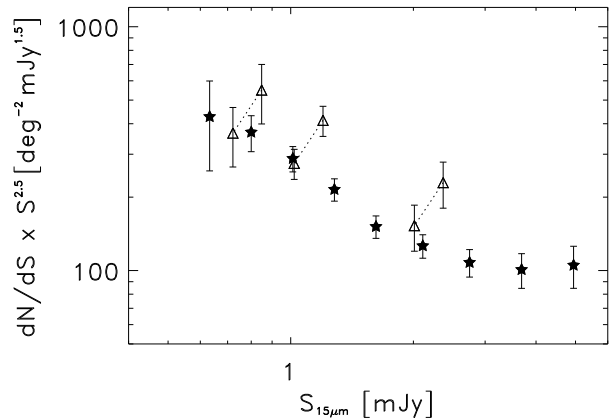,width=9cm}}
\caption{Zoom of the normalized differential source counts at
15 $\mu$m. The symbols are the
same as in figure \ref{fig_diffcount}: filled stars for ELAIS counts and open triangles for
Lockman Deep counts. The upper Lockman points are the original counts, while the lower ones
are the counts obtained with 15\% lowered fluxes.}
\label{lock_count}
\end{figure}

The reason for this discrepancy is still not completely understood. It
could be due to different separate causes or to a combination of them. A possible reason might be
attributed to different data reduction methods applied to different surveys. For example, our survey
has been reduced with the {\it Lari method} (see Paper I), while the Marano and the Lockman Deep Surveys have
been reduced with the {\it PRETI method} (Starck et al. 1999) and the Lockman Shallow Survey has been 
reduced with the {\it triple beam switch method} of IAS (Des\'ert et al. 1999). 
We must note that the {\it PRETI method} was especially designed to account for all the spurious
effects of ISOCAM data in a more complete way than the {\it triple beam switch method}.
A comparison between
these two methods performed in the HDF-N (Aussel et al. 1999a) has produced similar results, 
although not all the sources detected by one method were present in the list found by the other method. 
Moreover, flux densities derived for the common sources with the {\it triple beam switch method} were 
sistematically lower, by a factor of 0.82, than the {\it PRETI} fluxes. In fact, each method measures
fluxes in a different way (i.e. auto-simulations for {\it Lari}, photometry aperture plus implicit colour
correction for {\it PRETI}, fit with a fixed width Gaussian for the {\it triple beam switch}) and this
might produce small differences in the photometry of the objects. 
However, if our fluxes were on the same scale as the {\it PRETI} ones, the assumption that the 
{\it triple beam switch} fluxes are 20\% lower would furtherly increase the observed discrepancy
between our counts and the Lockman Shallow counts. Viceversa, good agreement between Lockman
Deep counts and our counts would be obtained if the Lockman Deep fluxes were systematically
higher than ours by about 15\% (see figure \ref{lock_count}). The flux calibration of our
catalogue, as described in detail in Paper I, has been tested using the stars in the field
and resulted on the same flux scale of IRAS (relative flux scale of 1.096 $\pm$ 0.044),
with photometric errors not larger than 10\%.

Another possible cause of the counts difference could be an incomplete star subtraction performed
in the Lockman Surveys, whose brightest flux density bins (at 2-5 mJy) might contain between 
20 and 50\% of stars (as shown in section \ref{counts}).

Finally, part of the observed discrepancy could also be due to cosmic variance
affecting small area fields, like the Marano Deep Field (0.2 sq. deg.) and the Lockman Deep and 
Shallow Fields (respectively 0.14 and 0.54 sq. deg.).

\subsection{Models and Interpretation}
\label{model}
As already mentioned, at flux densities $\gsimeq$ 1-2 mJy, the 
ELAIS counts are consistent with the
expectation of models assuming no evolution for extragalactic
sources, while they strongly depart from no evolution predictions
at fainter fluxes. The almost flat differential counts (normalised
to Euclidean) extending from the IRAS fluxes to 1-2 mJy, followed by
the sudden upturn below, seem to require strong evolution
for a single population rather than for the whole population of 
15 $\mu$m galaxies.\\
Due to the uncertainties existing in our counts at faint fluxes,
in this paper we do not pretend to construct an evolutionary model 
based on our result, however,
in order to interpret our data and the evolution they seem to 
require, we have compared 
our counts with recent evolution models for mid-infrared galaxies
found in literature.
In particular, we have compared our counts with the models of
Xu (2000) and Franceschini et al. (2001).
Neither model is able to reproduce the 
sharp departure feature from no-evolution predictions, or the 
low `plateau' between 1-2 and 100 mJy shown by our data.
The Xu (2000) model is able to fit the Deep/Ultra-Deep ISOCAM counts by
considering a rather extreme luminosity evolution (i.e. $L(z) =
L(0) \times (1+z)^{4.5}$) for the whole infrared population. However,
it largely over-predicts our counts above 0.8-1 mJy and its
departure from no evolution predictions is far too smooth to 
reproduce the sharp upturn shown by our data around 2 mJy. 
The model predictions of Franceschini et al. (2001) are somewhat
steeper than the Xu ones, but still overestimating and smoother than  
our counts, though considering a combination of luminosity and
density evolution for star-forming galaxies only. 
The local luminosity functions (LLFs) on which these models are based are 
different: the one considered by Xu (2000) has been derived using the
bivariate (15 $\mu$m vs. 60 $\mu$m luminosities) method,
from an IRAS sample selected at 60 $\mu$m and observed
by ISOCAM at 15 $\mu$m, while the one used by Franceschini et al. (2001)
is an adapted combination of the 12 $\mu$m LLF
from Fang et al. (1998) and the bivariate 15 - 60 $\mu$m, converted
to 12 $\mu$m, from Xu et al. (1998). Franceschini et al. (2001) have
also tried to disentangle the contributions of different populations.
Due to its larger flexibility in allowing to play with the different populations
and their evolutionary properties, we have based our analysis on the 
Franceschini et al. (2001) models, trying to find a good fit 
to our data by varying the LLF free parameters.\\
These models are able to reproduce the Deep/Ultra-Deep ISOCAM counts
by considering different evolutionary properties for three different 
populations: non-evolving normal spirals, strongly evolving starburst
plus Seyfert 2 galaxies and evolving AGN 1. The latter are assumed to
evolve in luminosity as $L(z) = L(0) \times (1+z)^3$ up to $z = 1.5$
and constant luminosity density at higher redshift. For the population of 
star-forming
and Seyfert 2 galaxies, in a $H_0 = 50$ km s$^{-1}$ Mpc$^{-1}$, $\Omega_m 
= 0.3$, $\Omega_{\Lambda} = 0.7$ Universe, the best-fit to Deep/Ultra-Deep 
ISOCAM counts is found by Franceschini et al. (2001) by considering
luminosity evolution $L(z) = L(0) \times (1+z)^{3.8}$ and density 
evolution $\rho(L[z],z) = \rho_0(L) \times (1+z)^4$ up to $z_{break}=0.8$,
and no additional evolution at $z > z_{break}$.\\
\begin{figure*}
\begin{minipage}{180mm}
\centerline{\psfig{figure=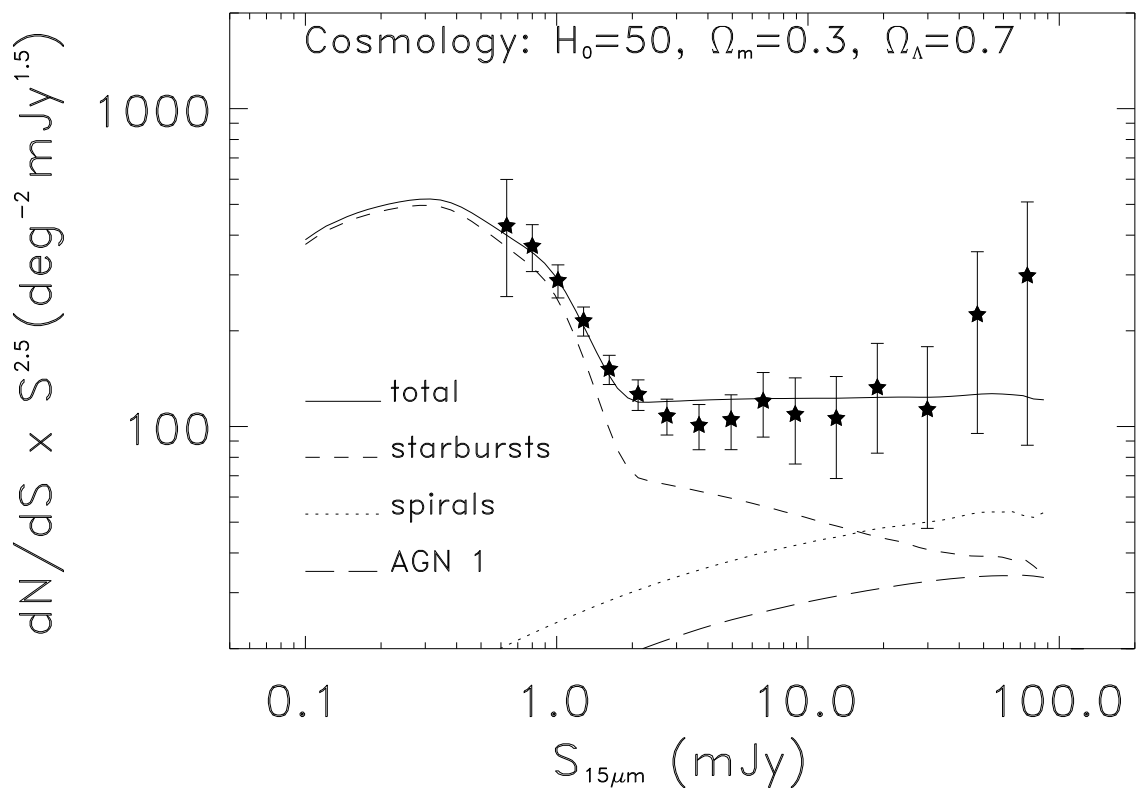,width=17cm}}
\caption{Best fit model (re-adapted from Franceschini et al.
2001) to ELAIS S1 source counts. The dotted line corresponds to the expected counts for
a population of non-evolving spirals. The short-dashed line is the modeled contribution
of a population of strongly evolving starburst (plus Seyfert 2) galaxies, while the
long-dashed line is the contribution of type I AGN. The solid line represents the expected
total source counts.}
\label{cnt_model}
\end{minipage}
\end{figure*}
We have not changed either the evolution (form and rate) for AGN 1, nor the 
evolutionary scheme for star-forming (plus Seyfert 2) galaxies. However, for the
latter, we have varied the luminosity and density evolution rates (respectively
$\alpha_L$ and $\alpha_D$ hereafter), $z_{break}$, and the LLF normalization.   
The crucial point is the star-forming galaxy LLF and in particular its shape at 
bright luminosities. In order to fit the sharp rise of our counts below 2 mJy,
we have found essential to introduce a luminosity break ($L_{break}$) in the 
star-forming LLF, above which it quickly drops to zero. 
This is also strongly required by the redshift
distribution for bright sources (above few mJy). In fact, without any luminosity 
break in the LLF, the redshift distribution predicted by the Franceschini et al.  
(2001) model for example for sources brighter than 5 mJy, shows a significant secondary
peak around $z \sim 1$, in addition to a low redshift peak. This is not consistent 
with the redshifts measured for bright ELAIS sources ($S \geq 2$ mJy; Gruppioni et 
al. 2001; La Franca, Gruppioni, Matute et al. 2002, in preparation), which are all found 
to be at rather low redshifts ($z < 0.3 - 0.4$). 

We have taken into account the redshift distribution constraints when 
looking for the best fit to our observed source counts. In particular, we have asked 
the model results to roughly agree with the following observational evidences:
\begin{itemize}
\item[1.] absence of high redshift peak for bright ($\geq 2$ mJy) sources 
(Gruppioni et al. 2001; La Franca, Gruppioni, Matute et al. 2002, in preparation);
\item[2.] majority of sources at moderate redshifts ($z < 0.3 - 0.4$) even at fluxes
$S \geq 0.8 - 1$ mJy, with a fraction of high $z$ sources not larger than 30 - 35\% 
(Pozzi, Ciliegi, Gruppioni et al., 2002, in preparation);  
\item[3.] redshift distribution for Deep surveys ($S \geq 0.1 - 0.2$ mJy) showing a
peak between $z = 0.5$ and $z = 1.2$ (HDF-North: Aussel et al. 1999b, Aussel et al. in
preparation, as reported by Franceschini et al. 2001; 
CFRS 11415+52: Flores et al. 1999). 
\end{itemize}

\begin{figure}
\centerline{\psfig{figure=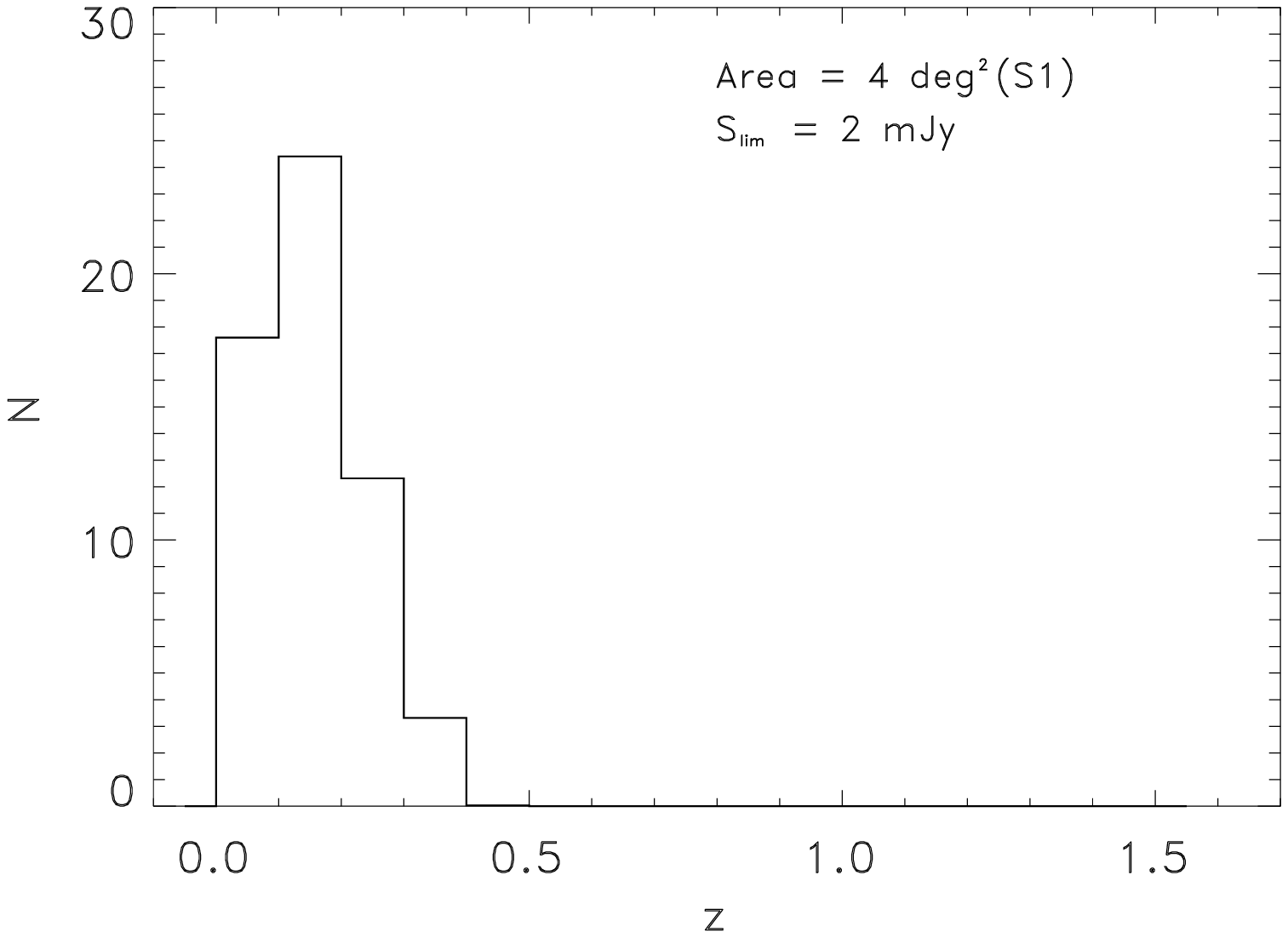,width=9cm}}
\centerline{\psfig{figure=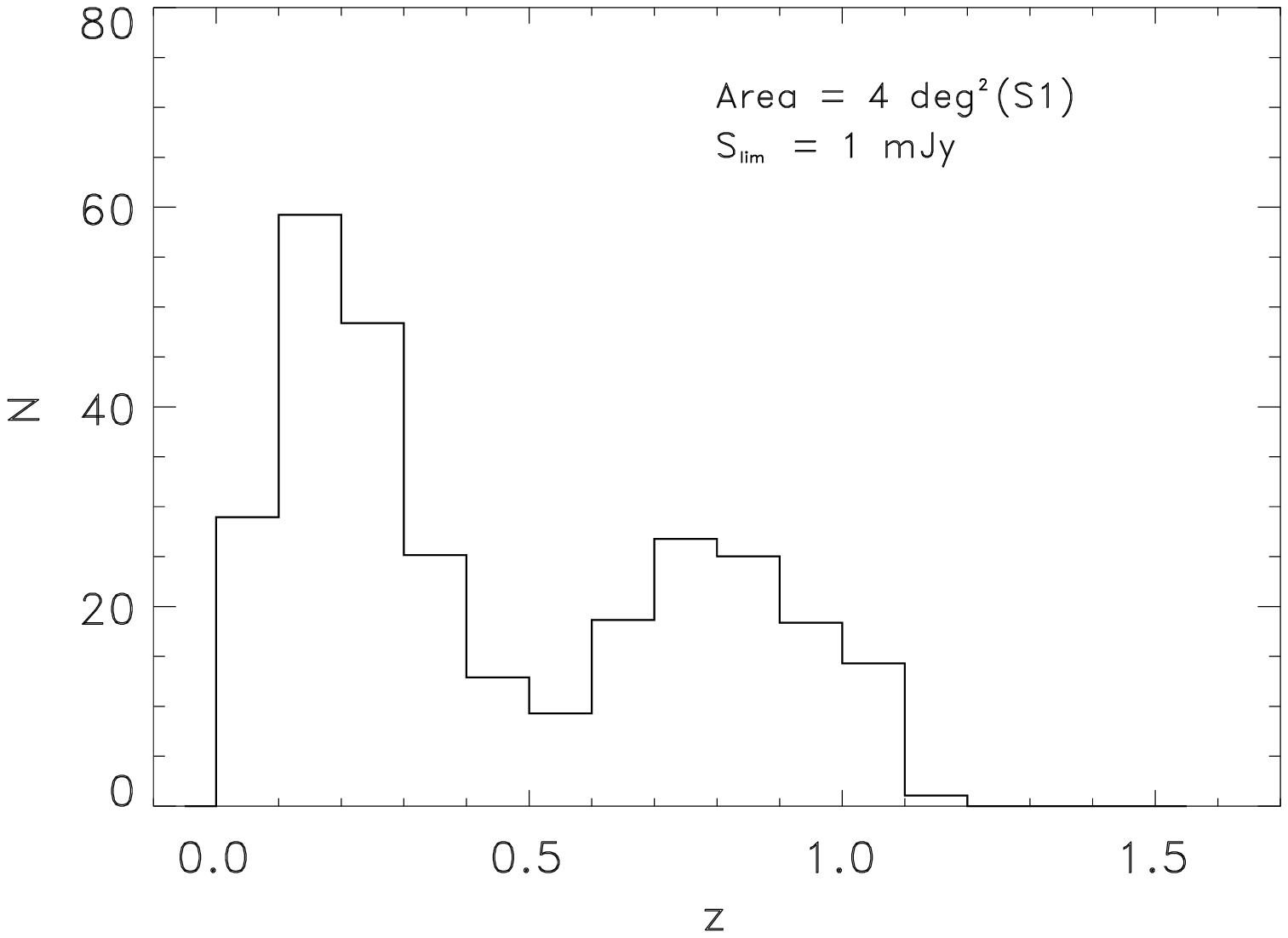,width=9cm}}
\centerline{\psfig{figure=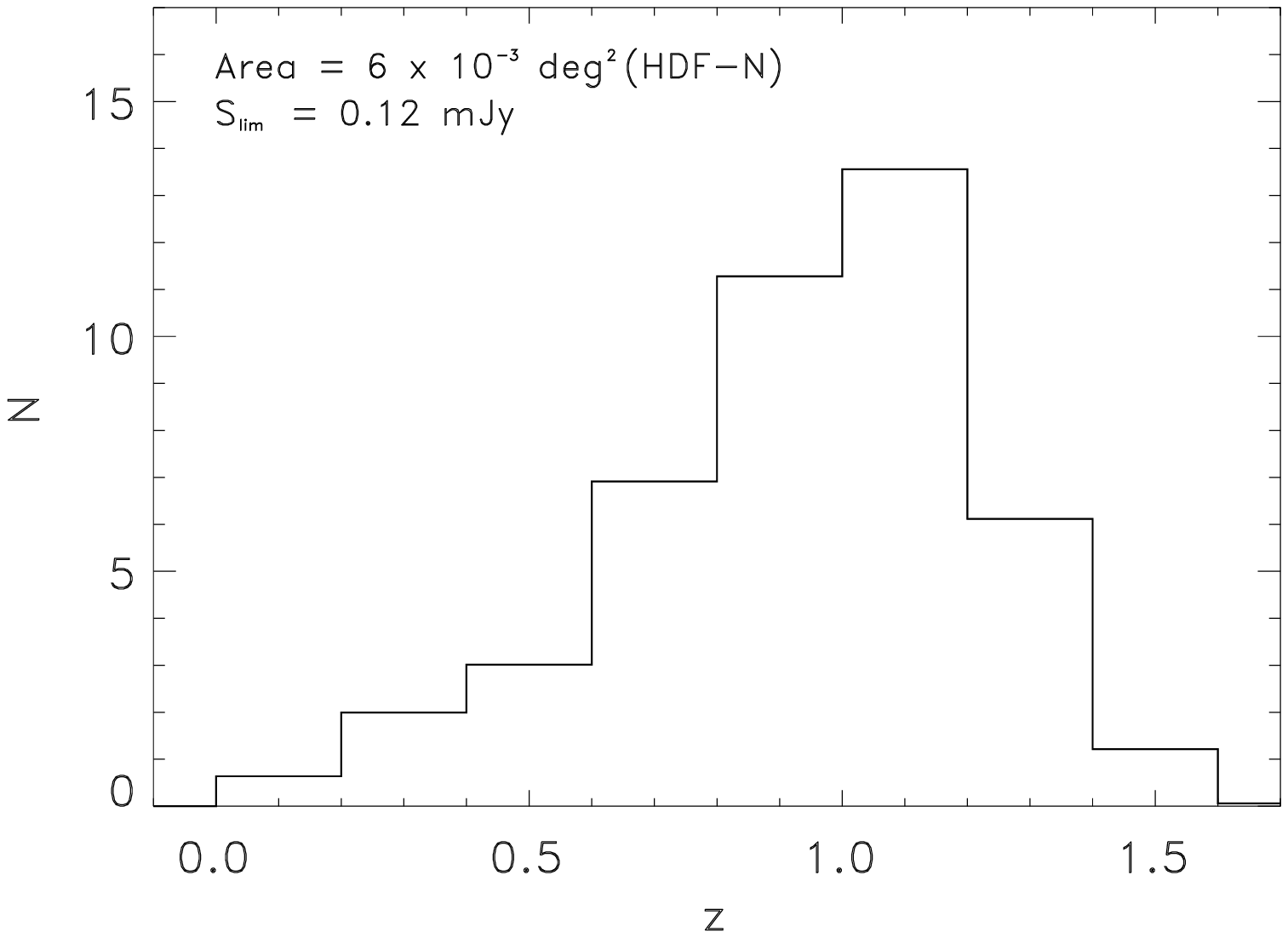,width=9cm}}
\caption{Redshift distributions predicted by our best fit model for 15 $\mu$m sources at different flux
levels and in different areas. $Top$: $S \geq 2$ mJy, area = 4 deg$^2$ (S1); $middle$:
$S \geq 1$ mJy, area = 4 deg$^2$ (S1); 
$bottom$: $S \geq 0.12$ mJy, area = $6 \times 10^{-3}$ deg$^2$ (HDF-N).
The latter is computed for the HDF-N area and to the same flux limit for a direct
comparison with the redshift distribution observed in that field (Aussel et al. 2002,
in preparation).}
\label{z_model}
\end{figure}

The best solution was found for $\alpha_L = 3.0$, $\alpha_D = 3.5$, $z_{break} = 1.1$,  
a LLF of evolving starburst population 40\% higher than the Franceschini
one and $L_{break} = 10^{10.8} L_{\odot}$. In figure \ref{cnt_model} the fit to source 
counts obtained with the above parameters is shown,
while the redshift distributions expected for our survey at $S_{15~\mu m} \geq$ 0.1, 1.0 
and 2.0 mJy are plotted in figure \ref{z_model}. 
These distributions are in good agreement with the preliminary results of optical
identification for the S1 sources on the DSS2. We find, in fact, that, while above 2 mJy
most sources have an optical counterpart brighter than $R \simeq 21.5$ (92 \% at $S_{15~\mu 
m}\geq 2$ mJy and 95\% at  $S_{15~\mu m} \geq 3$ mJy), between 1 and 2 mJy there is a 
quick drop in the
identification fraction (it goes down to $\sim$70\%), probably associated to a change
in population (i.e. appearance of the high-$z$ excess number of sources in the redshift
distribution). At $S_{15~\mu m} \geq 1$ mJy, the expected high-$z$ ($z \geq 0.6$) fraction 
of sources is $\sim$35\%, in good agreement with the fraction of S1 sources to the same flux
density without optical counterpart on the DSS2.  

Our result shows that significant evolution is needed for at least a class of extragalactic
objects, in order to explain the observed source counts below
a few mJy. Above about 10 mJy the counts are dominated by a non-evolving population
of normal spiral galaxies, while below this flux density a population of strongly evolving
starburst galaxies shows up and, rapidly rising, starts dominating the counts.
At fluxes $\lsimeq 3-5$ mJy, evolving starburst galaxies make up most of the observed
counts, being responsible for the peak around 0.3-0.4 mJy revealed by the Deep and
Ultra-Deep ISOCAM Surveys. The evolution required for this class of objects is lower
than found by Franceschini et al. (2001): $\alpha_L$ is 3.0 instead of 3.8 and $\alpha_D$
is 3.5 instead of 4.0. However, a turnover at higher $z$ ($z_{break} =$ 1.1 instead of 0.8) 
is needed in order to reproduce both the sharp rise of our counts and the faint flux peak of 
the Deep ISOCAM Surveys counts. This value for $z_{break}$ is intermediate between 0.8 found 
by Franceschini et al. (2001) and 1.5 found by Xu (2000), though the latter obtains a 
reasonably good fit to the Deep/Ultra-Deep counts and to
the redshift distribution of ISOCAM sources in the HDF-North (Aussel et al. 1999b), by 
considering pure luminosity evolution (with $\alpha_L = 4.5$) for the whole 15 $\mu$m 
extragalactic population. However, the Xu (2000) model is not able to fit the new
redshift distribution observed in the HDF-North with $>$ 90\% complete spectroscopic
identification, as derived by Aussel et al. (2002 in preparation) and reported by Franceschini
et al. (2001).
 
We have obtained an estimate of the 15 $\mu$m CIRB flux by directly integrating 
the best-fit model counts down to $S_{15~\mu m} = 50~ \mu$Jy: 2.2 nW m$^{-2}$ sr$^{-1}$.
This value is in good agreement with the computation done by Elbaz et al. (2001, as reported
by Chary \& Elbaz 2001), who finds
a value of 2.4 $\pm$ 0.5 nW m$^{-2}$ sr$^{-1}$ by integrating the observed ISOCAM
Deep/Ultra-Deep counts down to the same flux density limit. Our estimate of the
15 $\mu$m CIRB flux corresponds to about 67\% of the total resolved CIRB at 15 $\mu$m 
derived by Biviano et al. (2000) as 3.3 $\pm$ 1.3 nW m$^{-2}$ sr$^{-1}$.

\section{Conclusions}
\label{concl}
ISOCAM extragalactic source counts in the flux density range 0.5 -- 100 mJy 
have been derived for the ELAIS 15 $\mu$m samples obtained in the southern
hemisphere area with a new data reduction technique (see Paper I).
%The ELAIS counts are the only ones connecting the bright end of the ISOCAM
%Deep/Ultra-Deep counts to the fainter end of the IRAS counts.
Our counts sample very well the flux density region 
where Deep/Ultra-Deep ISOCAM counts start diverging from no evolution models. 
Indeed, we observe a significant change in slope from a value of $\sim$2.35
at fluxes higher than $S \simeq 2$ mJy to a very steep value ($\alpha \sim
3.60$) for fainter fluxes down to our flux limit ($S \simeq 0.5$ mJy).\\
This is in qualitative agreement with previous results, although the ELAIS counts 
show a somewhat steeper slope at faint fluxes than the other surveys 
($\alpha \simeq 3.0$ between 0.4 and 4 mJy; Elbaz et 
al. 1999b). At the faintest limit of our survey ($S \sim 0.5 - 1.0$ mJy),
where data from a number of other surveys exist, our counts agree with those
obtained in the Marano Deep Survey and are somewhat lower than those obtained
in other surveys. At brighter fluxes ($S \gsimeq 2 - 5$ mJy), where our data
are highly complete and statistically significant (because of the large sampled
area), our counts 
%are consistent with the ELAIS Preliminary Analysis (PA) counts
%of Serjeant et al. (2000), provided a downward correction of a factor 0.5 is applied 
%to the PA fluxes, but 
are significantly lower than the counts in the Lockman Deep
and Shallow Surveys. The observed difference might be attributable to different 
reduction methods applied to different surveys, to not complete star subtractions
and to cosmic variance that could affect small area surveys (i.e. Lockman Deep 
and Shallow, Marano Deep).

A good fit to our counts is obtained by re-adapting Franceschini et al. (2001) model
and introducing a luminosity break in the local luminosity function of the evolving 
population. Our solution considers no evolution for normal spiral galaxies (dominating
the counts at fluxes $\gsimeq 5 - 10$ mJy), a combination of luminosity and density
evolution ($L(z) = L(0) \times (1+z)^{3.0}$ and $\rho(L[z],z) = \rho_0(L) \times 
(1+z)^{3.5}$) up to $z_{break}=1.1$ for starburst galaxies, with a break in their LLF
at $L_{15~\mu m} = 10^{10.8} L_{\odot}$, and luminosity evolution ($L(z) = L(0) \times 
(1+z)^{3.0}$ up to $z=1.5$) for type 1 AGN. Strongly evolving starburst galaxies rise quickly
below $\sim 10$ mJy and start making up almost the totality of the observed counts at 
fluxes fainter than a few mJy. Our results are also in agreement with the observed
redshift distributions at different flux levels (from 10 down to 0.1 mJy), predicting 
a rather local population ($z_{med} \simeq 0.2$) of star-forming galaxies down to
$\sim$1.5 mJy and a rapid rise of a high-$z$ ($z_{med} \simeq 1$) excess of objects 
at fainter flux densities. This high-$z$ population totally dominates below 
$\sim 0.5$ mJy.

\section*{Acknowledgements}
This work was supported by the EC TMR Network programme FMRX--CT96--0068.
CG acknowledges partial support by the Italian Space Agency under
the contract ASI-I/R/27/00 and by the Italian Ministry for University and
Research (MURST) under grant COFIN99. The authors thank
D. Fadda for kindly providing the counts data relative to Deep and
Ultra-Deep ISOCAM Surveys.

\end{document}